\documentclass[11pt]{article}
\usepackage[dvips]{epsfig}
\usepackage[T1]{fontenc}
\usepackage[latin1]{inputenc}
\usepackage{graphicx}
\usepackage[english]{babel}
\usepackage{amsmath}
\usepackage{amssymb}
\usepackage{amsfonts}
\usepackage[T1]{fontenc}
\usepackage{color}
\usepackage{babel}
\usepackage{verbatim}
\usepackage[unicode=true,pdfusetitle,bookmarks=true,bookmarksnumbered=false,bookmarksopen=false,
 breaklinks=false,pdfborder={0 0 1},backref=false,colorlinks=true]{hyperref}
\hypersetup{linkcolor=blue,citecolor=blue}
\makeatletter
\usepackage[dvips]{epsfig}
\usepackage[T1]{fontenc}
\usepackage{color}
\usepackage{pdflscape}
\usepackage{cite}

\begin{document}

\title{\bf Generalized Vaidya Spacetime in Cotton and Conformal Killing  Theories}

\author{Metin G{\" u}rses $^a$\thanks{%
email: gurses@fen.bilkent.edu.tr},~Yaghoub Heydarzade $^a$\thanks{%
email: yheydarzade@bilkent.edu.tr}, and \c{C}etin \c{S}ent{\" u}rk $^b$\thanks{%
email: csenturk@thk.edu.tr}\\\\
\\{\small $^a$Department of Mathematics, Faculty of Sciences, Bilkent University, 06800 Ankara, Turkey}\\
{\small $^b$Department of Aeronautical Engineering,
University of Turkish Aeronautical Association, 06790 Ankara, Turkey}}
\date{\today}

\maketitle
\begin{abstract}
We demonstrate that the non-vacuum field equations of Cotton gravity and Conformal Killing gravity admit a generalized class of Vaidya-type solutions. In particular, beyond the standard induced term associated with the matter source, the generalized metric incorporates two additional correction terms of purely geometric origin, arising from the unique structure of these theories. This extended solution generalizes the classical Vaidya spacetime in General Relativity and offers new insights into the dynamics of radiating spacetimes within the framework of these  third-rank gravity theories.

\end{abstract}
\maketitle
\section{Introduction}
While General Relativity (GR) remains remarkably successful on solar system scales, it faces persistent challenges in accounting for galactic rotation curves, large-scale cosmic structure, and the observed late-time acceleration of the Universe. These issues have motivated the development of various modified gravity theories. Among these, Cotton Gravity (CG), proposed by Harada in 2021~\cite{haradacotton}, introduces a purely geometric extension of GR characterized
by third-rank tensorial field equations constructed from the Cotton tensor. Despite the higher-order nature of its field equations-originally involving third derivatives of the metric-CG has shown promise in describing galaxy dynamics without requiring dark matter~\cite{haradagalaxy}. Initial exact solutions, including Schwarzschild-like geometries, were presented in~\cite{haradacotton}, while more general spherically symmetric configurations were studied in~\cite{gog, lobo, heydarzade}. To simplify the technical complexity of CG, Mantica and Molinari proposed an equivalent second-rank formulation~\cite{mantica1}, in which the field equations are recast in terms of a Codazzi-type tensor, reducing the differential order and facilitating the construction of exact solutions. In this formulation, the field equations are given by
\begin{eqnarray}\label{cot}\nonumber
G_{\mu \nu}=\kappa T_{\mu \nu}+H_{\mu \nu}, \label{cot1} ~~~~~~\nabla_{\alpha}\,{\tilde H}_{\mu \nu}=\nabla_{\mu}\,{\tilde H}_{\alpha \nu},
\end{eqnarray}
where $\tilde H_{\mu \nu}=H_{\mu \nu}- \frac{1}{3}Hg_{\mu \nu}$ is a Codazzi tensor with $H=g^{\mu \nu}\,H_{\mu \nu}$, and $T_{\mu\nu}$ is the energy-momentum tensor of the external matter fields. This framework has facilitated the derivation of several non-trivial generalizations of GR solutions~\cite{susman}, although debates continue regarding the theory's internal consistency and physical viability~\cite{barg, haradareply, nuc1, sus1, nuc2, sus2}. A crucial distinction between CG and GR was highlighted in~\cite{gurses}, where it was shown that CG admits pp-wave solutions with non-flat wave surfaces, unlike GR. Moreover, AdS spherical and dS hyperbolic wave metrics in GR do not satisfy the field equations of CG. Additionally, CG has been found to possess identically vanishing conserved charges for all solutions in the absence of matter at the spatial boundary~\cite{tekin}, implying that vacuum solutions are degenerate with multi-black hole spacetimes, a striking departure from GR.  

After introducing Cotton Gravity (CG), Harada proposed another geometric theory known as Conformal Killing Gravity (CKG) in 2023, aimed at explaining the observed late-time accelerated expansion of the Universe without invoking dark energy~\cite{conf1, conf2}. Like CG, CKG features third-rank field equations. Shortly thereafter, Mantica and Molinari presented an equivalent second-rank formulation of this theory~\cite{mant}, in which the Einstein field equations are supplemented by a divergence-free conformal Killing tensor. The field equations in this formulation take the form
\begin{eqnarray}\label{con1}\nonumber
G_{\mu \nu}=\kappa T_{\mu \nu}+H_{\mu \nu}, ~~~~~~\nabla_{\alpha}\,{\tilde{H}}_{\mu \nu}+\nabla_{\mu}\,{\tilde{H}}_{\nu \alpha}+\nabla_{\nu}\,{\tilde{H}}_{\mu \alpha}=0, \label{conf1}
\end{eqnarray}
where $\tilde{H}_{\mu \nu}=H_{\mu \nu}-\frac{1}{6} H g_{\mu \nu}$ with $H=g^{\mu \nu}\,H_{\mu \nu}$, and $T_{\mu\nu}$ represents the matter energy-momentum tensor.
Spherically symmetric solutions in the context of CKG have been studied in~\cite{alan, mol, clem}. In addition, regular black hole and black bounce solutions have been reported in~\cite{lobo1, lobo2}. The Codazzi formulation of CKG and its cosmological applications have been explored in detail in~\cite{mant, manti, manti2}. Moreover, wave solutions in the CKG framework have been analyzed in~\cite{gurses, barnes}.

The Vaidya solution~\cite{Vaidya1951} is one of the most significant non-static solutions to Einstein's field equations, providing a natural dynamical generalization of the static Schwarzschild black hole. It is characterized by a time-dependent mass function $M(u)$, where $u$ is a null coordinate (retarded or advanced time), and an energy-momentum tensor describing an ingoing or outgoing null radiation flux. This solution serves as a classical model for a dynamical black hole undergoing evaporation or accretion, depending on the direction of the effective radiation flow \cite{evap1, evap3, evap4, evap6, parikh,
v4, v12}. Beyond modeling radiating black holes, the Vaidya spacetime has been widely used to study spherically symmetric gravitational collapse and has served as a valuable framework for testing the cosmic censorship conjecture~\cite{ccc,
naked1, naked3, kur}. It has also been proposed as a candidate for modeling the energy source for gamma-ray bursts, highlighting its astrophysical relevance~\cite{hark}. The original Vaidya solution has been generalized in multiple directions. It was  generalized to the charged case known as the Bonnor-Vaidya solution \cite{bonnor}.
A generalisation of the Vaidya solution constructed from type I and type II
energy-momentum tensors \cite{type} was introduced
in \cite{wanguli}. Similar  generalizations such as
 Vaidya-de Sitter \cite{mallet},
radiating dyon solution \cite{dyon}, Bonnor-Vaidya-de Sitter \cite{BVdS,
Patino, koberlin, saida},
the Husain solution \cite{husain}, and surrounded Vaidya and Bonnor-Vaidya
solutions \cite{yaghoub1, yaghoub2, yaghoub3} were also studied in the literature.

In this paper, we investigate Vaidya-type spacetimes within the frameworks of Cotton Gravity (CG) and Conformal Killing Gravity (CKG). The structure of the paper is as follows: In Section 2, we briefly introduce the generalized Vaidya metric and its key features. In Section 3, we derive a generalized Vaidya solution in the context of CG using the Codazzi formulation. Section 4 is devoted to obtaining a corresponding generalized solution within CKG. Finally, in Section 5, we present our concluding remarks. 
\section{Generalized Vaidya Metric}
We begin with considering the line element for a generalized Vaidya spacetime,
in Eddington-Finkelstein-like coordinates $x^\mu=(u,r,\theta,\phi)$, given
by
\begin{equation}\label{Vaidya}
ds^2=-f(u,r)\, du^2+2  du dr+r ^2 d \Omega^2,
\end{equation}
where  the metric function is defined as $f(u,r)=1-\frac{2 M(u,r)}{r}$, and
$d\Omega^2=d\theta^2+\sin^2d\phi^2$ represents the line element of the unit 2-sphere. This form generalizes the original Vaidya solution \cite{Vaidya1951} by allowing the mass function 
$M(u,r)$ to depend on both the null coordinate $u$ and the radial coordinate $r$.

The metric (\ref{Vaidya}) can be expressed in terms of the Kerr-Schild form \cite{KS,GG}
\begin{equation}\label{KSVaidya}
ds^2=ds_0^2+2Hdu^2~~\text{with}~~ds_0^2=-du^2+2  du dr+r ^2 d \Omega^2,
\end{equation}
where $ds_0^2$ is the flat background metric in advanced Eddington-Finkelstein coordinates and $H\equiv[1-f(u,r)]/2=M(u,r)/r$. As is well known, the Kerr-Schild type of metrics has proven itself as an extremely useful tool in getting exact solutions to the Einstein's        field equations in GR. Indeed, many exact solutions in GR can be brought into the Kerr-Schild form, see e.g. \cite{skmhh}. It has also recently been exploited to obtain exact wave-like solutions in CG and CKG theories \cite{gurses} and in more generic gravity theories \cite{gur1,gur2,gur3,gur4,gur5,gur6}. 

Although the Kerr-Schild form \eqref{KSVaidya} of the generalized Vaidya metric (\ref{Vaidya}) is nice and has very useful properties, it is technically more practical to use the original metric form \eqref{Vaidya} in our present analysis. So, we shall proceed with the metric given in (\ref{Vaidya}) from this point on and express it in terms of 
two null vectors $l_\mu$ and $n_\nu$ as 
\begin{equation}
g_{\mu\nu}=-l_\mu n_\nu-l_\nu n_\mu +h_{\mu\nu},
\end{equation}
where $h_{\mu\nu}=r^2\delta_\mu^\theta\delta_\nu^\theta+r^2\sin^2\theta
\delta_\mu^\phi\delta_\nu^\phi$ is the metric of the two-sphere of radius $r$ and the null vectors are defined
by
\begin{eqnarray}
l_\mu&=&\delta^0_\mu, ~~~~~~~~~~~~l^\mu~=\delta^\mu_1,\nonumber\\
n_\mu&=& \frac{f}{2}\delta^0_\mu -\delta^1_\mu, ~~~~n^\mu=-\delta^\mu_0 -\frac{f}{2}\delta^\mu_1.
\end{eqnarray}
These vectors satisfy the relations
\begin{equation}
l_\mu n^\mu = -1, \qquad l_\mu l^\mu = 0, \qquad n_\mu n^\mu = 0, \qquad l^\mu h_{\mu\nu}=0, \qquad n^\mu h_{\mu\nu}=0,
\end{equation}
confirming that both $l^\mu$ and $n^\mu$  are null and mutually normalized.
Moreover, the covariant derivatives of these vectors take the following forms
\begin{eqnarray}
&&\nabla_\mu l_\nu =-\frac{f^\prime}{2}l_\mu l_\nu +\frac{1}{r} h_{\mu\nu},\nonumber\\
&&\nabla_\mu n_\nu =\frac{f\,f^\prime}{4}l_\mu l_\nu
-\frac{f^\prime}{2}\delta^1_\nu l_\mu -\frac{f}{2r} h_{\mu\nu},\nonumber\\
&&l^{\mu}\, \nabla_{\mu}\, l_{\nu}=0,
\end{eqnarray}
where a prime denotes partial differentiation with respect to $r$. 

\section{Vaidya-Type Solutions in Cotton Gravity}
In this section, we explore the existence and structure of generalized Vaidya solutions in the framework of CG. The field equations of CG, in the Codazzi formulation, are given by
\begin{eqnarray}\label{CFEs}
G_{\mu \nu}&=&\kappa T_{\mu \nu}+H_{\mu \nu},\nonumber\\
\nabla_{\alpha}\, \tilde H_{\mu \nu}&=&\nabla_{\mu}\, \tilde H_{\alpha \nu},
\end{eqnarray}
where $\tilde H_{\mu \nu}=H_{\mu \nu}-\frac{1}{3}Hg_{\mu\nu}$ with $H=g^{\mu\nu}H_{\mu\nu}$, and $T_{\mu\nu}$ represents the energy-momentum tensor of the matter fields. The tensor $H_{\mu\nu}$ encapsulates the geometric modifications to the Einstein field equations induced by CG and includes contributions from third-order derivatives of the metric in its original formulation.

In order to analyze the existence of Vaidya-type solutions in CG, we consider the following distinct cases:
\begin{itemize}
\item Case I:  $M=M(u),~~ T_{\mu\nu}=0$,
\item Case II:  $M=M(u,r),~~ T_{\mu\nu}=0$,
\item Case III:  $M=M(u),~~ T_{\mu\nu}(u)\neq0$,
\item Case IV:  $M=M(u),~~ T_{\mu\nu}(u,r)\neq0$,
\item Case V:   $M=M(u,r),~~ T_{\mu\nu}(u,r)\neq0$.
\end{itemize}
Each case corresponds to a physically distinct scenario, ranging from general radiating systems with matter sources to purely vacuum configurations. The case V represents the most general non-static setup, while the remaining cases are its specific subclasses with progressively more restrictive assumptions.
In our analysis, the energy-momentum tensor is taken to be a combination of null and timelike matter components. This allows us to model a wide range of non-vacuum configurations, incorporating both radiative and non-radiative sources. We begin by deriving the most general solution corresponding to the Case V. Subsequently, we examine the special cases listed above, identifying the precise conditions under which a generalized Vaidya metric solves the CG field equations.

In the most general case (Case V) of a Vaidya-type configuration in CG with a non-vanishing energy-momentum tensor, following~\cite{wanguli, husain}, we decompose the total energy-momentum tensor as
\begin{equation}\label{Tmunu}
T_{\mu\nu}=T^{(n)}_{\mu\nu}+T^{(m)}_{\mu\nu}, 
\end{equation}
where
\begin{eqnarray}\label{emt}
&&T^{(n)}_{\mu\nu}=\mu l_\mu l_\nu,\label{Tn}\\
&&T^{(m)}_{\mu\nu}=(\rho+p)(l_\mu n_\nu+l_\nu n_\mu)+pg_{\mu\nu}.\label{Tm}
\end{eqnarray}
Here, $T^{(n)}_{\mu\nu}$ represents the energy-momentum tensor of a null fluid with energy density $\mu(u,r)$ propagating along null hypersurfaces $u = \text{const.}$, while $T^{(m)}_{\mu\nu}$ corresponds to a timelike fluid with energy density $\rho(u,r)$ and pressure $p(u,r)$. The total energy-momentum tensor (\ref{Tmunu}) can be written in the matrix form as
 \begin{equation}
T_{\mu \nu }(u,r)\text{ = }\left[
\begin{array}{cccc}
 \left(1-\frac{2 M (u, r) }{r}\right)\rho(u,r)+\mu(u,r) & -\rho(u, r)  & 0 & 0 \\
 -\rho(u,r)  & 0 & 0 & 0 \\
 0 & 0 & r^2p(u,r) & 0 \\
 0 & 0 & 0 & r^2 \sin ^2\theta p(u,r) \\
\end{array}
\right].
\end{equation}
Using this expression, the non-vanishing components of the modified geometric tensor $\tilde{H}_{\mu\nu} = H_{\mu\nu} - \dfrac{1}{3} H g_{\mu\nu}$ are calculated to be
 \begin{eqnarray}
\tilde H_{00} &=& \frac{1}{3r^3}\left\{2(r-2M)(M'-rM'')+6r\dot{M}-\kappa r^2[(r-2M)(2p+\rho)+3r\mu]\right\}, \\
\tilde H_{01} &=& -\frac{1}{3r^2}\left[2(M'-rM'')-\kappa r^2(2p+\rho)\right], \\
\tilde H_{22} &=& \frac{1}{3} \left[4M'-rM''-\kappa r^2(p+2\rho)\right], \\
\tilde H_{33} &=& \sin^2\theta\,  \tilde H_{22}.
\end{eqnarray}
Substituting these into the CG field equations (\ref{CFEs}) leads to the following system of differential equations:
\begin{eqnarray}
F_{001} &\equiv & \frac{1}{3r^4}\biggl\{2r(6\dot{M}-4r\dot{M}'+r^2\dot{M}'')+2(r-2M)(2M'-2rM''+r^2M''')\nonumber\\
&&~~~~~~~~~~~~~~~~~~~~~~~~~~~~~~~~~~~~~~~~~+\kappa r^3\left[(r-2M)(2p'+\rho')+r(2\dot{p}+\dot{\rho})+3r\mu'\right]\biggr\}=0,\label{F001}\\
F_{011} &\equiv & -\frac{1}{3r^3}\left[2(2M'-2rM''+r^2M''')+\kappa r^3(2p'+\rho')\right]=0, \label{F011}\\
F_{022} &\equiv & -\frac{1}{3r}\left\{6\dot{M}-4r\dot{M}'+r^2\dot{M}''+\kappa r^2[r(\dot{p}+2\dot{\rho})-3\mu]\right\}=0, \label{F022}\\
F_{122} &\equiv & -\frac{1}{3r}\left\{2M'-2rM''+r^2M'''+\kappa r^2[3(p+\rho)+r(p'+2\rho')]\right\}=0, \label{F122}\\
F_{033} &\equiv & \sin ^2\theta F_{022}=0, \\
F_{133} &\equiv & \sin ^2\theta F_{122}=0,\label{F133}
\end{eqnarray}
where we have made use of the definition
\begin{equation}\label{FtensorCG}
F_{\alpha\mu\nu}\equiv\nabla_{\alpha}\, \tilde{H}_{\mu \beta}-\nabla_{\mu}\, \tilde{H}_{\beta \alpha}.
\end{equation}
Multiplying (\ref{F011}) by $r^2/2$ and subtracting the resulting equation from (\ref{F122}), we obtain 
\begin{equation}\label{rhoEq}
r\rho'+2(p+\rho)=0.
\end{equation}
Assuming an equation of state of the form $p = w\rho$, where $w$ is a constant, Eq.~(\ref{rhoEq}) can be integrated to give
\begin{equation}\label{rhour}
\rho(u,r)=\rho_0(u)r^{-2(1+w)}.
\end{equation}
Substituting this result, along with $p = w\rho$, back into Eq.~(\ref{F011}) and solving for $M(u, r)$, we obtain
\begin{equation}\label{Mur}
M(u,r)=C_0(u)+\frac{1}{2}C_1(u) r^2+ \frac{1}{3}C_2(u)r^3+\frac{\kappa \rho_0(u)}{2(1-2w)}r^{1-2w}.
\end{equation} 
With this mass function, Eq.~(\ref{F022}) gives the null energy density as
\begin{equation}\label{muur}
\mu(u,r)=\frac{\dot{C_0}(u)}{\kappa r^2}+\frac{\dot{\rho_0}(u)}{(1-2w)}r^{-(1+2w)}.
\end{equation}
All the other equations are satisfied, and thus, the functions given in Eqs.~(\ref{rhour})-(\ref{muur}) define a complete family of exact solutions to the CG field equations. These include both null and timelike matter fields and exhibit generalized Vaidya-like behavior.

We have the following remarks.
\\
\noindent
\textbf{Remark 1:} For $w = -1/2$ and $w = -1$, the term with $r^{1 - 2w}$ in Eq.~(\ref{Mur}) becomes proportional to $r^2$ and $r^3$, respectively. In these special cases, the corresponding contribution can be absorbed into redefinitions of $C_1(u)$ and $C_2(u)$, preserving the general structure of the mass function:  
\begin{eqnarray}
&&M(u,r)=C_0(u)+\frac{1}{2}\tilde{C}_1(u) r^2+ \frac{1}{3}C_2(u)r^3,~~~~\text{for $w=-\frac{1}{2}$},\\
&&M(u,r)=C_0(u)+\frac{1}{2}C_1(u) r^2+ \frac{1}{3}\tilde{C}_2(u)r^3,~~~~\text{for $w=-1$}.
\end{eqnarray}
\noindent
\textbf{Remark 2:}  From the general solution given in Eqs.~(\ref{rhour})-(\ref{muur}), it follows that by choosing $\rho_0(u) = 0$, the energy density and pressure of the timelike matter component vanish, i.e., $\rho(u,r) = p(u,r) = 0$. In this case, the contribution $T^{(m)}_{\mu\nu}$ in Eq.~(\ref{Tmunu}) drops out, and the total energy-momentum tensor reduces to a purely null component:
 \begin{equation}
T_{\mu\nu} = T^{(n)}_{\mu\nu} = \mu(u,r)\, l_\mu l_\nu, \qquad \mu(u,r) = \frac{\dot{C}_0(u)}{\kappa r^2}.
\end{equation}
The corresponding Vaidya-type metric remains a solution of CG with the mass function given by
\begin{equation}
 M(u,r) = C_0(u) + \frac{1}{2}C_1(u)\, r^2 + \frac{1}{3}C_2(u)\, r^3.
\end{equation}
This illustrates that even in the absence of timelike matter fields, CG admits radiating Vaidya-type spacetimes sourced purely by null radiation, generalizing the standard Vaidya solution within this extended gravitational framework.\\
\\
\noindent
\textbf{Remark 3:} In Case I, $M=M(u),~~ T_{\mu\nu}=0$,
the field equations  \eqref{F001}-\eqref{F133} simplify to $\dot{M}(u) = 0$, hence
\begin{equation}
 \quad M(u) = \text{const.},
\end{equation}
which can also be seen from the general solution given in Eqs.~(\ref{rhour})-(\ref{muur}) by taking $\rho_0=0$, $C_0=const.$, $C_1=0$, and $C_2=0$. This means that in the vacuum configuration of CG with $ M = M(u)$, the mass function must be constant. As the result, the generalized  solution reduces to the standard Schwarzschild spacetime. This implies that the original Vaidya solution cannot be a vacuum solution of CG through the $H_{\mu\nu}$ tensor.\\
\\
\noindent
\textbf{Remark 4:} In Case II, $M=M(u,r),~~ T_{\mu\nu}=0$,  the mass function takes the form
\begin{equation}
M(u,r)=C_0+\frac{1}{2}C_1(u) r^2+ \frac{1}{3}C_2(u)r^3,
\end{equation}
where $C_0$ is constant. This implies that unlike Case I with $M = M(u)$, here the inclusion of $r$-dependence in the mass function allows non-trivial dynamical
solutions even in the absence of explicit matter fields. The resulting spacetime describes a novel dynamical solution sourced purely by geometric corrections from CG through the $H_{\mu\nu}$ tensor.  This solution may be interpreted
as the Schwarzschild black hole in a dynamical de Sitter-like background. 
\\
\\
\textbf{Remark 5:} In Case III, $M=M(u),~~ T_{\mu\nu}(u)\neq0$,
the field equations \eqref{F001}-\eqref{F133} give
\begin{equation}
p=-\rho=const.=c, ~~~~M(u)=const., ~~~~ \mu=0.
\end{equation}
This solution can also be obtained from the general solution given in Eqs.~(\ref{rhour})-(\ref{muur}) when one takes $w=-1$, $\rho_0=const.$, $C_0=const.$, $C_1=0$, and $C_2=0$. In this case the non-vanishing energy-momentum tensor takes the form
\begin{equation}
T_{\mu \nu }\text{ = }\left[
\begin{array}{cccc}
 -c\left(1-\frac{2 M}{r}\right) & c  & 0 & 0 \\
 c  & 0 & 0 & 0 \\
 0 & 0 & cr^2 & 0 \\
 0 & 0 & 0 & cr^2 \sin^2\theta \\
\end{array}
\right].
\end{equation}
This case shows that CG permits static solutions sourced by a combination of timelike and null fluids only under very specific constraints: the pressure and energy density must be constant, equal in magnitude but opposite in sign, and the null fluid must vanish. Moreover, the mass function becomes constant, preventing radiating solutions in this configuration.  In the limit \( c = 0 \), this solution reduces to the vacuum case, i.e.
Case I.
\\
\\
\noindent
\textbf{Remark 6:} In Case IV, $M=M(u),~~ T_{\mu\nu}(u,r)\neq0$,
the field equations \eqref{F001}-\eqref{F133} together with the equation of state $p=w\rho$ admit consistent solutions for only the specific values $w=-1$ and $w=-\frac{1}{2}$. In these cases, the matter fields and mass function take the form
\begin{equation}
p=w\rho=-\rho_0(u),~~~ ~~~~M(u)=C_0(u), ~~~~ \mu(u,r)=\frac{2\dot C_0(u)}{\kappa r^2}+\frac{\dot \rho_0(u)}{3}r.
\end{equation}
and 
\begin{equation}
p=w\rho=-\frac{\rho_0(u)}{2r},~~~ ~~~~M(u)=C_0(u), ~~~~ \mu(u,r)=\frac{2\dot C_0(u)}{\kappa r^2}+\frac{\dot \rho_0(u)}{2}.
\end{equation}
Hence, CG admits the original Vaidya solution in the presence of a  non-zero timelike matter source
in addition to the null matter source. This is an interesting difference of
this theory with respect to GR.  In the limit $\rho_0(u) = 0$, these solutions reduce to the original Vaidya solution supported only by the null matter source.

\vspace{0.5cm}
\noindent
\section{Vaidya-type Solutions in Conformal Killing Theory}
In this section, we investigate the existence of generalized Vaidya-type solutions in the framework of Conformal Killing Gravity (CKG). The field equations of CKG are given by
\begin{eqnarray}\label{CKGEqs}
&&G_{\mu \nu}=\kappa T_{\mu \nu}+H_{\mu \nu},\nonumber\\
&&\nabla_{\alpha}\, \tilde{H}_{\mu \beta}+\nabla_{\mu}\, \tilde{H}_{\beta \alpha}+\nabla_{\beta}\, \tilde{H}_{\alpha \mu}=0,
\end{eqnarray}
where $\tilde{H}_{\mu \nu} = H_{\mu \nu} - \frac{1}{6} H g_{\mu \nu}$ with $H = g^{\mu \nu} H_{\mu \nu}$, and $T_{\mu \nu}$ is the matter energy-momentum tensor. The tensor $H_{\mu \nu}$ encapsulates the higher-order (third-derivative) modifications to Einstein's field equations, characteristic of the CKG framework.

We assume the most general case, Case V, where the metric function is $M = M(u, r)$, and the energy-momentum tensor is given by \eqref{Tmunu} with  $\mu=\mu(u,r)$, $\rho=\rho(u,r)$, and $p=p(u,r)$. The non-vanishing components of the  $\tilde{H}_{\mu \nu}$ tensor are computed as 
\begin{eqnarray}
\tilde H_{00} &=& \frac{1}{3r^3}\left\{(r-2M)(4M'-rM'')+6r\dot{M}-\kappa r^2[(r-2M)(p+2\rho)+3r\mu]\right\},\\
\tilde H_{01} &=& -\frac{1}{3r^2}\left[4M'-rM''-\kappa r^2(p+2\rho)\right], \\
\tilde H_{22} &=& \frac{1}{3} \left[M'-rM''-\kappa r^2(2p+\rho)\right], \\
\tilde H_{33} &=& \sin^2\theta\, \tilde H_{22}.
\end{eqnarray}
Hence, the CKG equations (\ref{CKGEqs}) give
\begin{eqnarray}
F_{001} &\equiv & -\frac{1}{3r^4}\biggl\{2r(6\dot{M}+r\dot{M}'-r^2\dot{M}'')+(r-2M)(8M'-5rM''+r^2M''')\nonumber\\
&&~~~~~~~~~~~~~~~~~~~~~~~~~~~~~~~~~~~~~~~~~+\kappa r^3\left[(r-2M)(p'+2\rho')-2r(\dot{p}+2\dot{\rho})+3r\mu'\right]\biggr\}=0,\label{F001CKG}\\
F_{011} &\equiv & \frac{1}{3r^3}\left[2(8M'-5rM''+r^2M''')+2\kappa r^3(p'+2\rho')\right]=0, \label{F011CKG}\\
F_{022} &\equiv & \frac{1}{3r}\left\{2(6\dot{M}+r\dot{M}'-r^2\dot{M}'')-\kappa r^2[r(2\dot{p}+\dot{\rho})+6\mu]\right\}=0, \label{F022CKG} \\
F_{122} &\equiv & -\frac{1}{3r}\left\{2(8M'-5rM''+r^2M''')-\kappa r^2[6(p+\rho)-r(2p'+\rho')]\right\}=0, \label{F122CKG}\\
F_{033} &\equiv &\sin ^2\theta\,  F_{022}=0, \\
F_{133} &\equiv & \sin ^2\theta\, F_{122}=0,\label{F133CKG}
\end{eqnarray}
where 
\begin{equation}\label{FtensorCKG}
F_{\alpha\mu\nu}\equiv\nabla_{\alpha}\, \tilde{H}_{\mu \beta}+\nabla_{\mu}\, \tilde{H}_{\beta \alpha}+\nabla_{\beta}\, \tilde{H}_{\alpha \mu}.
\end{equation}
Multiplying (\ref{F011CKG}) by $r^2$ and adding the resulting equation to (\ref{F122CKG}) yield
\begin{equation}
r\rho'+2(p+\rho)=0,
\end{equation}
which is the same equation we obtained before in (\ref{rhoEq}) in CG. Hence,
similarly, assuming a barotropic equation of state $p = w\rho$ with constant $w$, this equation can be integrated as
\begin{equation}\label{rhour1}
\rho(u,r)=\rho_0(u)r^{-2(1+w)}.
\end{equation}
Inserting this solution, along with $p=w\rho$, back into (\ref{F011CKG}) and solving the resulting equation for $M$ produces
\begin{equation}\label{Mur1}
M(u,r)=C_0(u)+\frac{1}{3}C_1(u) r^3+ \frac{1}{5}C_2(u)r^5+\frac{\kappa \rho_0(u)}{2(1-2w)}r^{1-2w}.
\end{equation} 
From (\ref{F022CKG}), the null energy density $\mu(u, r)$ is then determined as
\begin{equation}\label{muur1}
\mu(u,r)=\frac{1}{\kappa r^2}\left(2\dot{C_0}(u)+\frac{1}{3}\dot{C_1}(u)r^3-\frac{3}{5}\dot{C_2}(u)r^5\right)+\frac{\dot{\rho_0}(u)}{(1-2w)}r^{-(1+2w)}.
\end{equation}
However, substituting the equations (\ref{rhour1}), (\ref{Mur1}), and (\ref{muur1}), together with $p=w\rho$, back into the full field equations  (\ref{F001CKG})-(\ref{F133CKG}), shows that consistency is achieved only when
\begin{equation}
C_1 = \text{const.}, \quad C_2 = \text{const.}
\end{equation}
Thus, the final general form of the mass function and the null fluid energy density in CKG are
\begin{eqnarray}\label{Mur2}
&&M(u,r)=C_0(u)+\frac{1}{3}C_1 r^3+ \frac{1}{5}C_2r^5+\frac{\kappa \rho_0(u)}{2(1-2w)}r^{1-2w},\label{Mur22}\\
&&\mu(u,r)=\frac{2\dot{C_0}(u)}{\kappa r^2}+\frac{\dot{\rho_0}(u)}{(1-2w)}r^{-(1+2w)}\label{Mur33}.
\end{eqnarray}
We have the following remarks on this solution.
\\
\\
\noindent
\textbf{Remark 1:} As in CG, the special cases $w=-1/2$ and $w=-1$ in the last term of (\ref{Mur2}) can be absorbed into the second and third terms by defining new functions $C_1(u)$ and $C_2(u)$, respectively.\\
\\
\noindent
\textbf{Remark 2:}   From the general solution presented in Eqs.~(\ref{rhour1}), (\ref{Mur22}), and (\ref{Mur33}), it follows that setting $\rho_0(u) = 0$ eliminates the timelike matter component, yielding $\rho(u,r) = p(u,r) = 0$. Consequently, the term $T^{(m)}_{\mu\nu}$ in Eq.~(\ref{Tmunu}) vanishes, and the total energy-momentum tensor reduces to a purely null form
\begin{equation}
T_{\mu\nu} = T^{(n)}_{\mu\nu} = \mu(u,r)\, l_\mu l_\nu, \qquad \mu(u,r) = \frac{2\dot{C}_0(u)}{\kappa r^2}.
\end{equation}
In this scenario, the Vaidya-type metric remains a valid solution to the field equations of CKG, with the corresponding mass function given by
\begin{equation}
M(u,r) = C_0(u) + \frac{1}{3}C_1\, r^3 + \frac{1}{5}C_5\, r^5.
\end{equation}
Notably, the appearance of quadratic and cubic powers in $r$ within the mass function absent in standard GR is a direct consequence of the higher-order geometric corrections inherent to CKG. These additional terms reflect the influence of conformal Killing constraints on the spacetime geometry and signal modifications to the usual radiating solutions.
This demonstrates that even in the absence of timelike matter fields, CKG admits nontrivial radiating solutions sourced purely by null radiation. In particular, the theory naturally generalizes the standard Vaidya spacetime to include de Sitter-like corrections, revealing a key structural difference from GR.
\\
\\
\noindent
\textbf{Remark 3:} In Case I, corresponding to $M = M(u)$ with $T_{\mu\nu} = 0$, the CKG field equations simplify considerably and lead directly to the condition
\begin{equation}
M(u) = \text{const}.
\end{equation}
This can also be obtained from the general solution given in (\ref{rhour1}), (\ref{Mur22}), and (\ref{Mur33}) by taking $\rho_0=0$, $C_0=const.$, $C_1=0$, and $C_2=0$. This result indicates that in a vacuum configuration of CKG with $M = M(u)$, the mass function must remain constant. Consequently, the generalized Vaidya metric reduces to the standard Schwarzschild spacetime. Therefore, no radiating Vaidya-like solutions are permitted in vacuum within CKG. \\
\\
\noindent
\textbf{Remark 4:} In Case II, $M=M(u,r),~~ T_{\mu \nu}=0$, the
consistency of the field equations demands the static mass function of the form
\begin{equation}
M(u,r)=C_0+\frac{1}{3}C_1 r^3+ \frac{1}{5}C_2r^5.
\end{equation}
This represents that in a vacuum configuration of CKG with $M = M(u,r)$, the mass function must remain constant. Here, the generalized Vaidya metric reduces to a generalized Schwarzschild spacetime. Therefore, no radiating Vaidya-like solutions are permitted in vacuum within CKG.
One notes
that this is in contrast to  CG possessing a dynamical solution in this case.
\\
\\
\textbf{Remark 5:} In Case III, $M=M(u),~~ T_{\mu \nu}(u)\neq0$,
the field equations \eqref{F001CKG}-(\ref{F133CKG}) give
\begin{equation}
p=-\rho=const.=c, ~~~~M(u)=C_0, ~~~~ \mu=
0,
\end{equation}
which can also be obtained from the general solution given in (\ref{rhour1}), (\ref{Mur22}), and (\ref{Mur33}) if one takes $w=-1$, $\rho_0=const.$, $C_0=const.$, $C_1=0$, and $C_2=0$. This is the static Schwarzschild solution in the presence of a non-vanishing energy-momentum
source. This implies another difference of CKG and GR.
\\
\\
\noindent
\textbf{Remark 6:} In Case IV, $M=M(u),~~ T_{\mu \nu}(u,r)\neq0$,
the field equations \eqref{F001CKG}-(\ref{F133CKG}) together with the equation of state $p=w\rho$ admit consistent solutions for only the specific values $w=-1$ and $w=-2$. In these cases, the matter fields and mass function take the form
\begin{equation}
p=w\rho=-\rho_0=const.,~~~ ~~~~M(u)=C_0(u), ~~~~ \mu(u,r)=\frac{2\dot C_0(u)}{\kappa r^2},
\end{equation}
and 
\begin{equation}
p=w\rho=-2\rho_0 r^2,~~~ ~~~~M(u)=C_0(u), ~~~~ \mu(u,r)=\frac{2\dot C_0(u)}{\kappa r^2}.
\end{equation}
Thus, CKG also admits the original Vaidya solution in the presence of a  non-zero timelike constant matter source
in addition to the null matter source.  In the limit $\rho_0= 0$, both the solutions above reduce to the original Vaidya solution supported only by the null matter source.
\section{Conclusion}

In this work, we have studied the generalized Vaidya spacetime within the framework of CG and CKG theories. This spacetime generalizes the original Vaidya solution by the inclusion of the energy-momentum tensor of a timelike matter field in addition to that of a null fluid. Assuming the barotropic equation of state $p=w\rho$ with $w=const.$ for the timelike matter field, we obtained exact solutions to the third-order equations of motion of CG and CKG. In both theories, in the most general case, the timelike energy density $\rho(u,r)$ has the same solution which is time-dependent and exhibiting a power-law behavior controlled by the constant $w$; i.e.,
\begin{equation}\nonumber
\rho(u,r)=\rho_0(u)r^{-2(1+w)}.
\end{equation}
Interestingly this energy density contributes to the mass function $M(u,r)$ and the null energy density $\mu(u,r)$ in the same fashion in both CG and CKG. In fact, in CG, they are obtained as 
\begin{eqnarray}
&&M(u,r)=C_0(u)+\frac{1}{2}C_1(u) r^2+ \frac{1}{3}C_2(u)r^3+\frac{\kappa \rho_0(u)}{2(1-2w)}r^{1-2w},\nonumber\\
&&\mu(u,r)=\frac{\dot{C_0}(u)}{\kappa r^2}+\frac{\dot{\rho_0}(u)}{(1-2w)}r^{-(1+2w)},\nonumber
\end{eqnarray}
while, in CKG, they are 
\begin{eqnarray}
&&M(u,r)=C_0(u)+\frac{1}{3}C_1 r^3+ \frac{1}{5}C_2r^5+\frac{\kappa \rho_0(u)}{2(1-2w)}r^{1-2w},\nonumber\\
&&\mu(u,r)=\frac{\dot{2C_0}(u)}{\kappa r^2}+\frac{\dot{\rho_0}(u)}{(1-2w)}r^{-(1+2w)}.\nonumber
\end{eqnarray}
These form interesting exact solutions to the field equations of CG and CKG, and special cases can be obtained by appropriately setting the functions/constants appearing in these solutions. Especially, it is worth noting that, in both theories, the functions/constants $C_1$ and $C_2$ control the contributions to the mass function $M(u,r)$'s coming purely from geometry, not related to the energy-momentum tensor. In fact, when there is no matter at all; i.e, $T_{\mu\nu}=0$, or equivalently $\rho(u,r)=\mu(u,r)=0$, the functions $C_0$'s become constants in the mass functions of both CG and CKG, and this still makes the solution time-dependent in CG through the functions $C_1$ and $C_2$, which might be the geometric generalization of the Vaidya solution in the
standard GR requiring null matter for its very existence, while being time-independent in CKG.

\end{document}